\documentclass{PoS}

\title{New Results from the T2K Experiment: Observation of $\nu_e$ Appearance in a $\nu_\mu$ Beam}

\ShortTitle{Observation of $\nu_e$ Appearance from a $\nu_\mu$ Beam}

\author{\speaker{Michael Wilking} on behalf of the T2K collaboration\\
        TRIUMF\\
        E-mail: \email{wilking@triumf.ca}}

\abstract{The T2K Experiment has observed the appearance of electron
  neutrinos in a muon-neutrino beam. Twenty eight neutrino candidates
  pass the selection cuts with a predicted background of 4.64$\pm$0.53
  events, which corresponds to a 6.8$\sigma$ exclusion of
  $\theta_{13}=0$. When comparing the $\theta_{13}=0$ hypothesis to
  the best fit of the full electron momentum and angle distribution,
  $\theta_{13}=0$ is excluded at 7.5$\sigma$ for
  $\sin^22\theta_{23}=1$ and $\delta_{CP}=0$. In addition, results
  from the T2K $\nu_\mu$ disappearance analysis have also been updated
  to improve the treatment of full 3-flavor oscillations.}

\FullConference{The European Physical Society Conference on High Energy Physics -EPS-HEP2013\\
		18-24 July 2013\\
		Stockholm, Sweden}

\begin{document}


The phenomenon of neutrino oscillation has been well established
experimentally. However, the explicit transformation of one neutrino
flavor to another particular flavor has never before been conclusively
observed. Using a dataset corresponding to 6.39$\times 10^{20}$
protons on target (POT), the T2K experiment has made the first
observation of neutrino appearance via the $\nu_\mu\to\nu_e$
oscillation channel.

The T2K neutrino beam is generated using a 30~GeV proton beam produced
at J-PARC. Charged pions and kaons are produced in interactions of the
beam with a graphite target that is embedded within the first of three
magnetic focusing horns, and are then allowed to decay in a 100~m
drift volume to produce the neutrino beam. The results presented in
this update were made possible due to the efforts of the J-PARC
accelerator division and the operators of the neutrino beamline. In
2013, stable running at 220 kW was achieved, including a world-record
1.2$\times 10^{14}$ protons per pulse, which allowed T2K to more than
double the 3.01$\times 10^{20}$ POT used in the previous T2K $\nu_e$
appearance measurement~\cite{t2koldnue}.

T2K measures the unoscillated neutrino flux at a near detector complex
(ND280) located 280~m downstream of the neutrino production target.
Since the previous T2K electron appearance
measurement~\cite{t2koldnue}, the calibration, reconstruction, Monte
Carlo simulation (MC), and analysis of the near detector data have all
been significantly improved.  Many of these improvements affect the
matching of tracks found in the ND280 time projection chambers (TPCs)
to hits in the Fine-Grained Detectors (FGDs), which serve as the
neutrino targets.
If the matching terminates prematurely for an external track passing
through the FGD, the track will be incorrectly reconstructed as having
originated within the FGD. This out-of-fiducial-volume background was
the largest detector systematic uncertainty in many of the muon
momentum bins in the previous analysis.

To mitigate these issues, a more precise treatment of the detector
material was implemented, as well as an improved treatment of
high-angle tracks. Improvements were made to the TPC reconstruction to
reduce the sensitivity to delta rays, which can induce spurious
rotations in the reconstructed TPC trajectories. Improvements to the
reconstruction to handle variations in the TPC drift velocity were
introduced, and the monitoring and stability of the drift velocity
were enhanced. These improvements reduced the systematic uncertainty
associated with out-of-fiducial-volume events from the 1-3\% range,
depending on the muon momentum, to below 1\% for all momenta above 400
MeV/c. Below 400 MeV/c, these events remain the dominant detector
systematic uncertainty, but the size of the error is reduced relative
to the previous analysis. Above 400 MeV/c, the largest detector
uncertainty is due to uncertainties in pion hadronic interaction cross
sections.

The $\nu_{\mu}$ charged-current (CC) interactions used to constrain
the neutrino energy spectrum and cross section parameters are selected
using the same criteria as the previous T2K
analysis~\cite{t2koldnue}. The muon is identified as the
highest-momentum, negative-curvature track that emerges from the FGD
fiducial volume and has an energy deposit in the TPC immediately
downstream of the FGD that is consistent with a muon. Tracks found in
the TPC upstream of the FGD are used to veto external backgrounds.

The previous analysis divided the CC sample into a "CCQE-like" sample
and a "CCnonQE-like" sample. The CCQE-like sample was the subset of
the CC sample which contained only a single TPC-FGD track and no late
time energy deposit in FGD1 consistent with an electron from a
$\pi^+\to\mu^+\to e^+$ decay chain. All other CC events were placed in
the CCnonQE-like sample. These two samples were then simultaneously
fit to the data. With the addition of the Run 3 data, this previous
near detector analysis was limited by systematic uncertainties. Run 3
increased the POT for which near detector data were collected from
0.96$\times 10^{20}$ to 2.56$\times 10^{20}$. However, these additional
data only provided a reduction in the uncertainty in the far detector
event rate prediction from 5.7\% to 4.7\% for sin$^22\theta_{13}$ of
0.1. In particular, the fit had limited ability to constrain the cross
section parameters associated with CC$\pi^+$ interactions, since the
CCnonQE-like sample had a CC$\pi^+$ purity of only 29\%.

To improve the near detector constraint, the current analysis now
incorporates three samples: CC0$\pi$, CC1$\pi^+$, and CCother. These
samples are defined based on the number of pions found in the final
state. A $\pi^+$ can be identified in one of three ways: a TPC2+FGD1
track with a TPC charge deposition consistent with a pion, a late-time
energy deposit consistent with a decay electron in FGD1, or a track
fully contained within FGD1 with an FGD charge deposition consistent
with a pion. To tag a $\pi^-$, only negative curvature TPC2+FGD1
tracks are used, since fully-contained FGD1 tracks do not allow for
sign selection. A $\pi^0$ is identified if there exists a track in
TPC2 with a charge deposition consistent with an electron. Events
containing no pions are placed in the CC0$\pi$ sample, events with
exactly one $\pi^+$ and no $\pi^-$ or $\pi^0$ are classified as
CC1$\pi^+$, and all other CC events are placed in the CCother
sample. With these criteria, the resulting purities of the CC0$\pi$,
CC1$\pi^+$, and CCother samples are 72.6\%, 49.4\%, and 73.8\%,
respectively.

\begin{figure}[hb!]
 \begin{center}
 \includegraphics[width=0.45\textwidth]{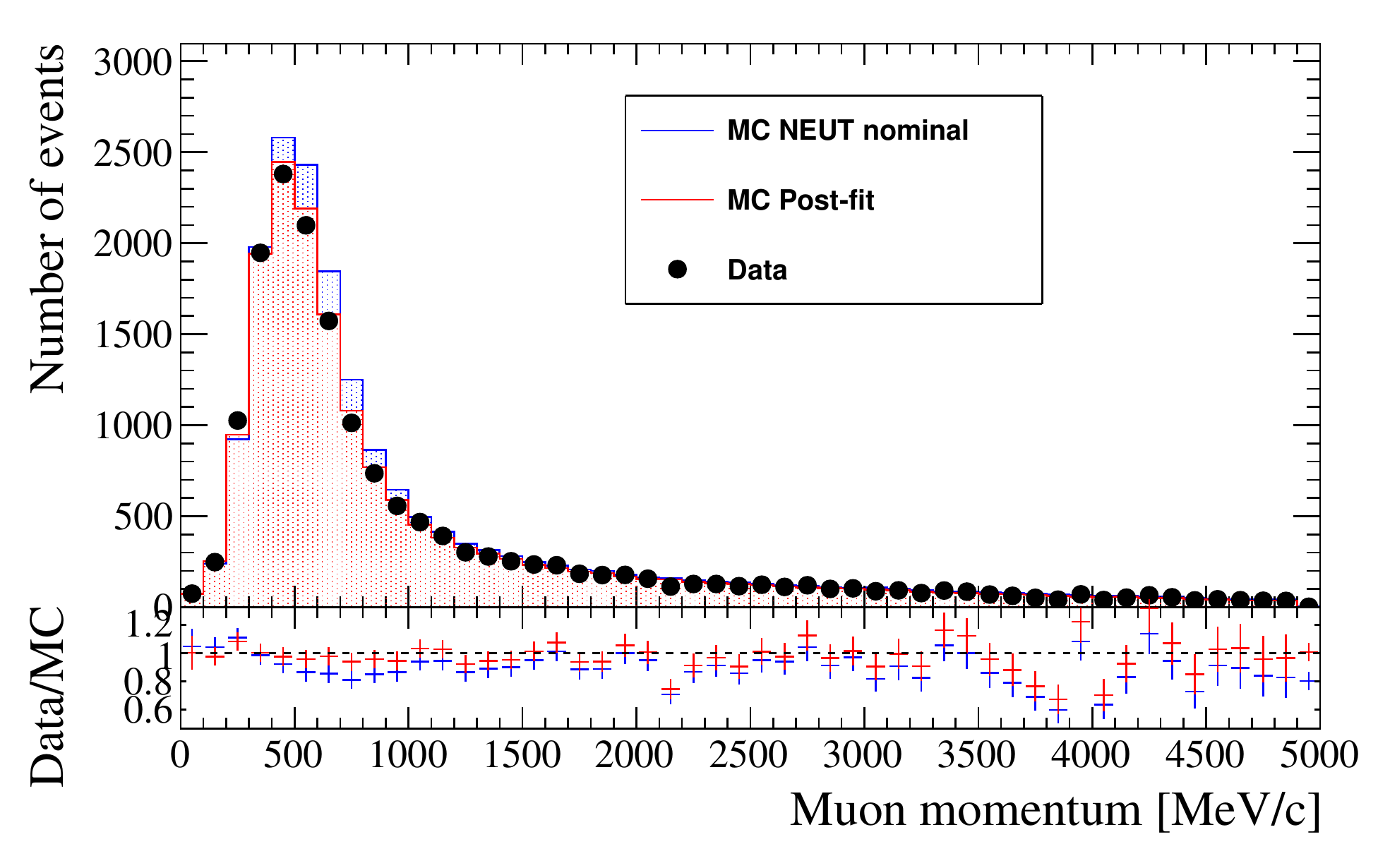} 
 \includegraphics[width=0.45\textwidth]{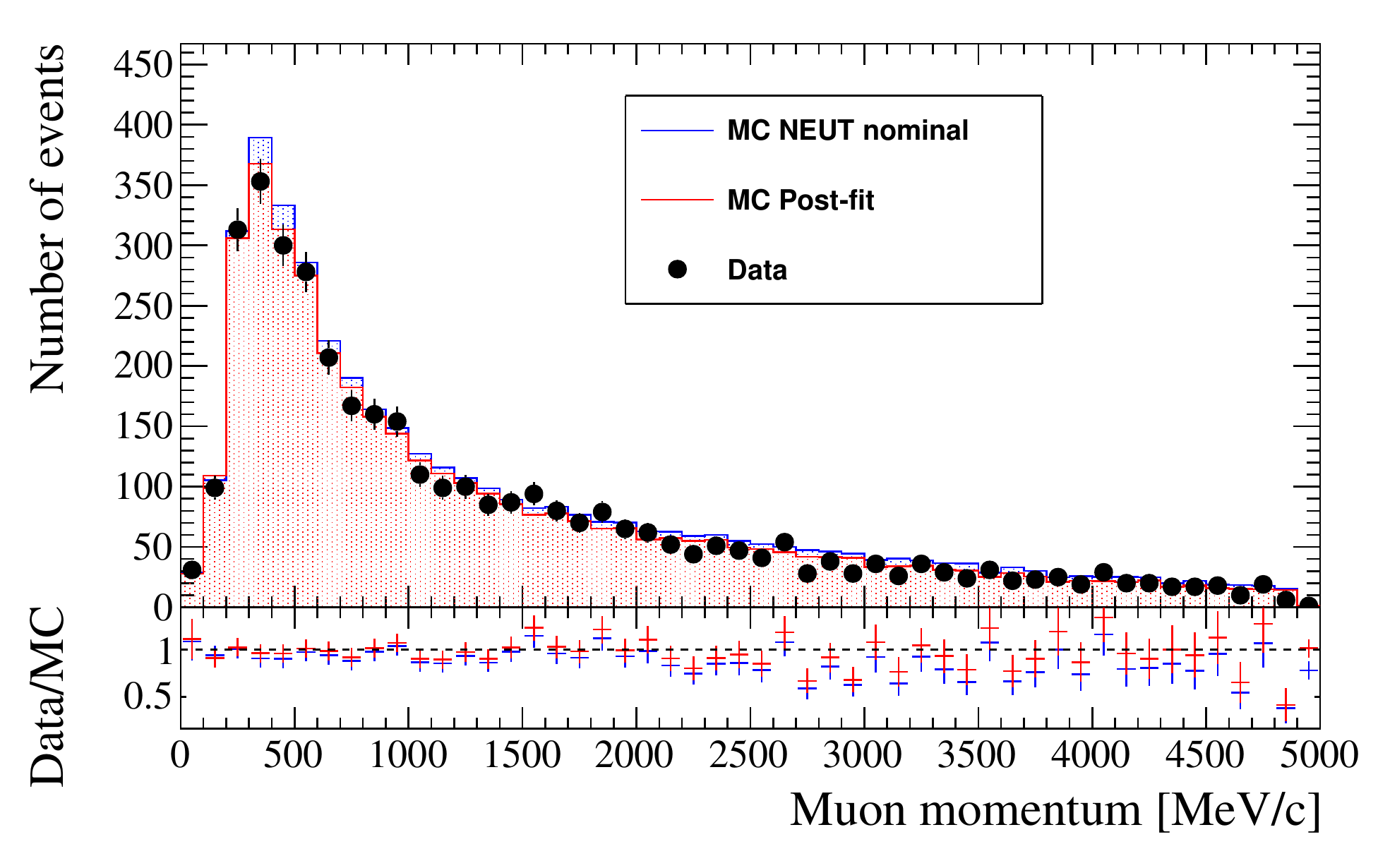}
 \includegraphics[width=0.45\textwidth]{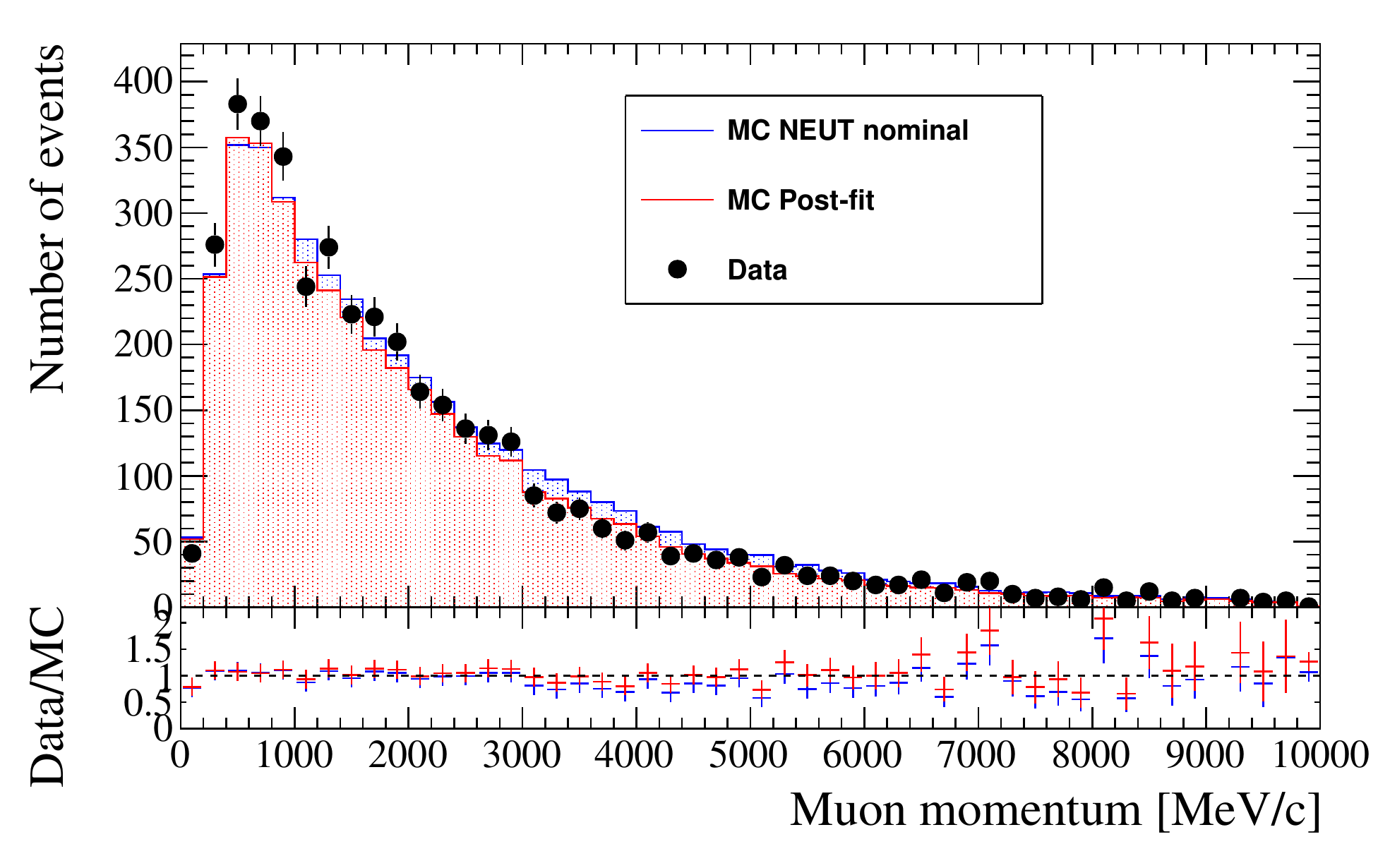} 
 \end{center}
 \caption{The muon momentum distributions of the CC0$\pi$ (top-left),
   CC1$\pi^+$ (top-right), and CCother (bottom) samples are shown. The
   data (black points) are compared to the MC before (blue histogram)
   and after (red histogram) the near detector constraint is
   applied.}  \label{fig:nd280mumom}
\end{figure}

When the near detector constraint is extracted using the same Run 1-3
data as the previous analysis, but with the new CC0$\pi$, CC1$\pi^+$,
and CCother samples, the uncertainty on the far detector event rate is
reduced from 4.7\% to 3.5\% for sin$^22\theta_{13}$ of 0.1,
sin$^22\theta_{23}$ of 1.0, and $\delta_{CP}$ of 0. Adding the Run 4
data further reduces the far detector event rate uncertainty to
3.0\%. The momentum distributions for each of the three event samples
before and after the near detector constraint is applied are shown in
Figure~\ref{fig:nd280mumom}.

The analysis of the data from the T2K far detector, Super-Kamiokande,
has also been improved. A new method for reconstructing particle
kinematics has been used for the first time in this analysis. This new
algorithm is a maximum likelihood fit, based on the mathematical formalism
developed by MiniBooNE~\cite{miniboonerecon}, that calculates charge and time
probability density functions (PDFs) for every photomulitplier tube
(PMT) for any given choice of the initial particle parameters. These
PDFs are calculated by integrating the light contributed by each
segment of the particle trajectory. Particle identification is
achieved by separately performing a fit for several different particle
hypotheses, and then comparing the resulting best-fit likelihood
values. Multiple-particle fit hypotheses, such as the two-photon
$\pi^0$ fit, are constructed by combining the charge and time PDFs for
each particle in the final state. Combined charge PDFs are created by
summing the charge contributions from each particle, and the combined
time PDFs are the weighted average of the individual particle PDFs
based on the distance to the PMT. By using the same likelihood
framework for both single- and multiple-particle final states, the
resulting fit likelihoods can be directly compared to determine the
final event topology.

Most of the $\nu_e$ event selection continues to use the
previously-existing Super-K reconstruction, and is unchanged relative
to the previous T2K result~\cite{t2koldnue}. Events with entering or
exiting particles are vetoed by requiring less than 16 hits in all hit
clusters in the outer-detector, and the reconstructed vertex in the
inner-detector is required to be at least 2~m from the nearest
wall. For the $\nu_e$ selection, only events with a single,
electron-like ring with an energy greater than 100~MeV are accepted,
and events with late-time signals consistent with a stopped muon or
pion decay are rejected to further reduce $\nu_\mu$-induced
background.

\begin{figure}[h!]
 \begin{center}
 \includegraphics[width=0.45\textwidth]{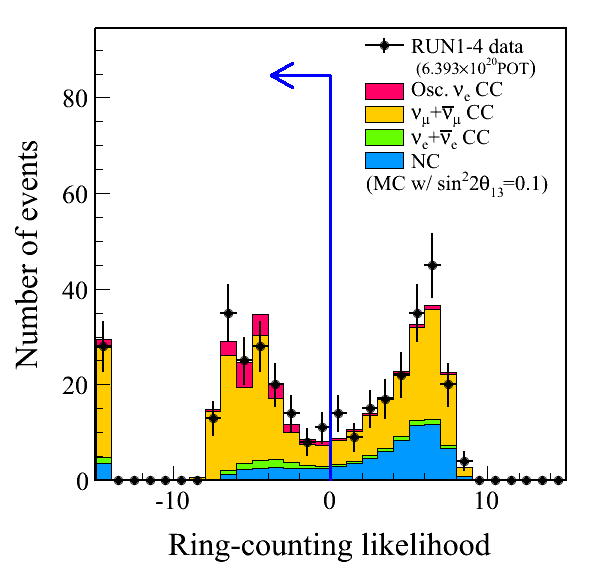}
 \includegraphics[width=0.45\textwidth]{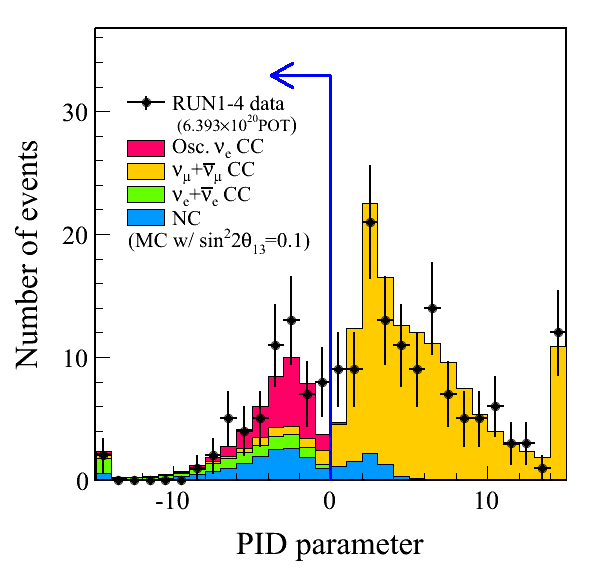}
 \end{center}
 \caption{The ring-counting (left) and electron/muon particle
   identification (right) distributions are shown for both data and
   MC. The vertical line indicates the selection cut in each
   distribution.}  \label{fig:eventselection}
\end{figure}

\begin{figure}[h!]

 \begin{center}
 \hspace{3mm}
 \includegraphics[width=0.45\textwidth]{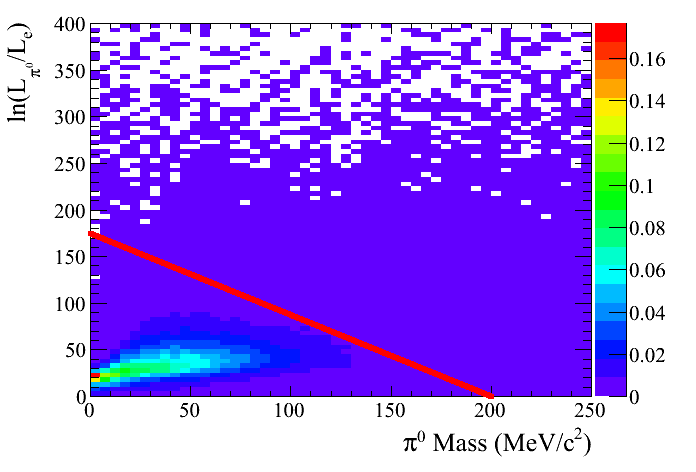} 
 \includegraphics[width=0.45\textwidth]{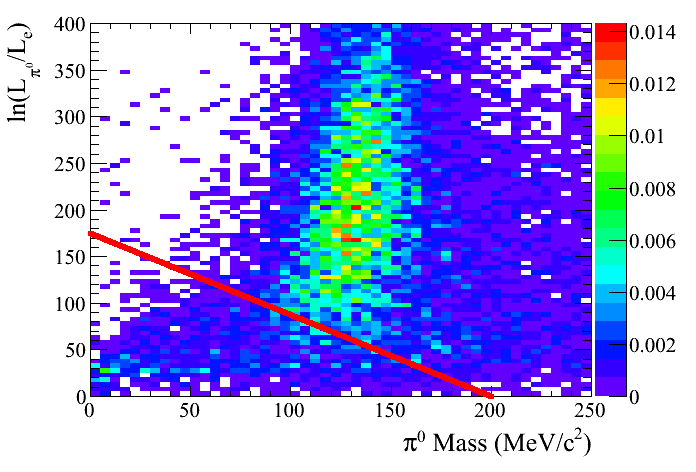}
 \includegraphics[width=0.45\textwidth]{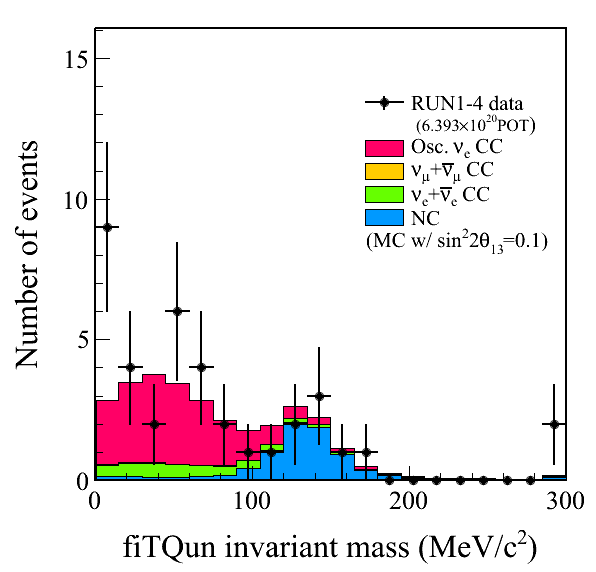} 
 \includegraphics[width=0.45\textwidth]{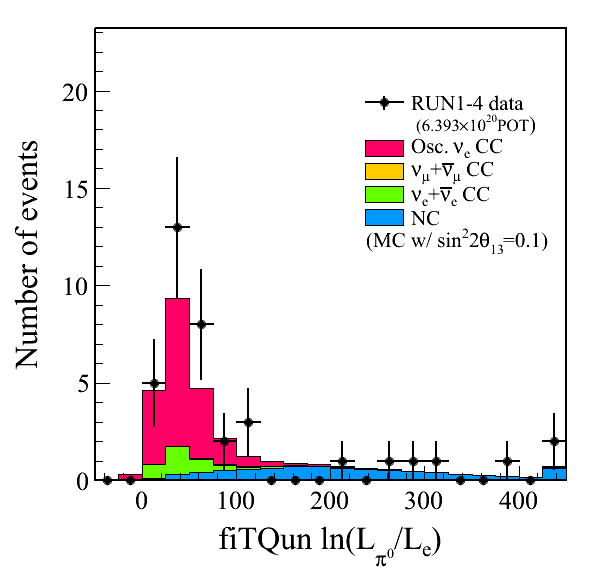}
 \end{center}
 \caption{The MC $\ln(L_{\pi^0}/L_{e})$ vs $\pi^0$ mass distributions
   from the new reconstruction algorithm (called fiTQun) are shown for
   both signal CC-$\nu_e$ (top-left) and background events that
   contain a $\pi^0$ (top-right). The bottom plots show comparisons
   between the data and MC in the reconstructed $\pi^0$ mass
   (bottom-left) and log-likelihood ratio
   (bottom-right).} \label{fig:pi0cut}
\end{figure}

The final step in the event selection now uses the new reconstruction
algorithm to remove additional $\pi^0$ background. The cut uses two
variables: the reconstructed two-photon mass from the $\pi^0$ fit, and
the ratio of the best-fit likelihoods from the $\pi^0$ fit and the
single-ring electron-hypothesis fit, $\ln(L_{\pi^0}/L_{e})$. The
distributions of both the signal CC-$\nu_e$ and the background
containing a $\pi^0$ are shown in Figure~\ref{fig:pi0cut}. A linear
cut in this two-dimensional space, defined as
$\ln(L_{\pi^0}/L_{e})<175-0.875 \times m_{\pi^0}$~(MeV/c$^{2}$), is
used to select the signal. This cut removes 70\% of the $\pi^0$
background that was allowed in the previous analysis with only a 2\%
decrease in the signal efficiency.  This improvement to the background
rejections improves the sensitivity for rejecting $\theta_{13}=0$ from
5.0$\sigma$ to 5.5$\sigma$. The final event sample contains 28 events
in the data with a predicted background of $4.64\pm 0.53$ events and
an expected signal sample of $20.4\pm 1.8$ for
sin$^22\theta_{13}=0.1$, sin$^22\theta_{23}=1$, $\delta_{CP}=0$, and
normal neutrino mass hierarchy.

\begin{figure}[h!]
 \begin{center}
 \hspace{5mm}
 \includegraphics[width=0.36\textwidth]{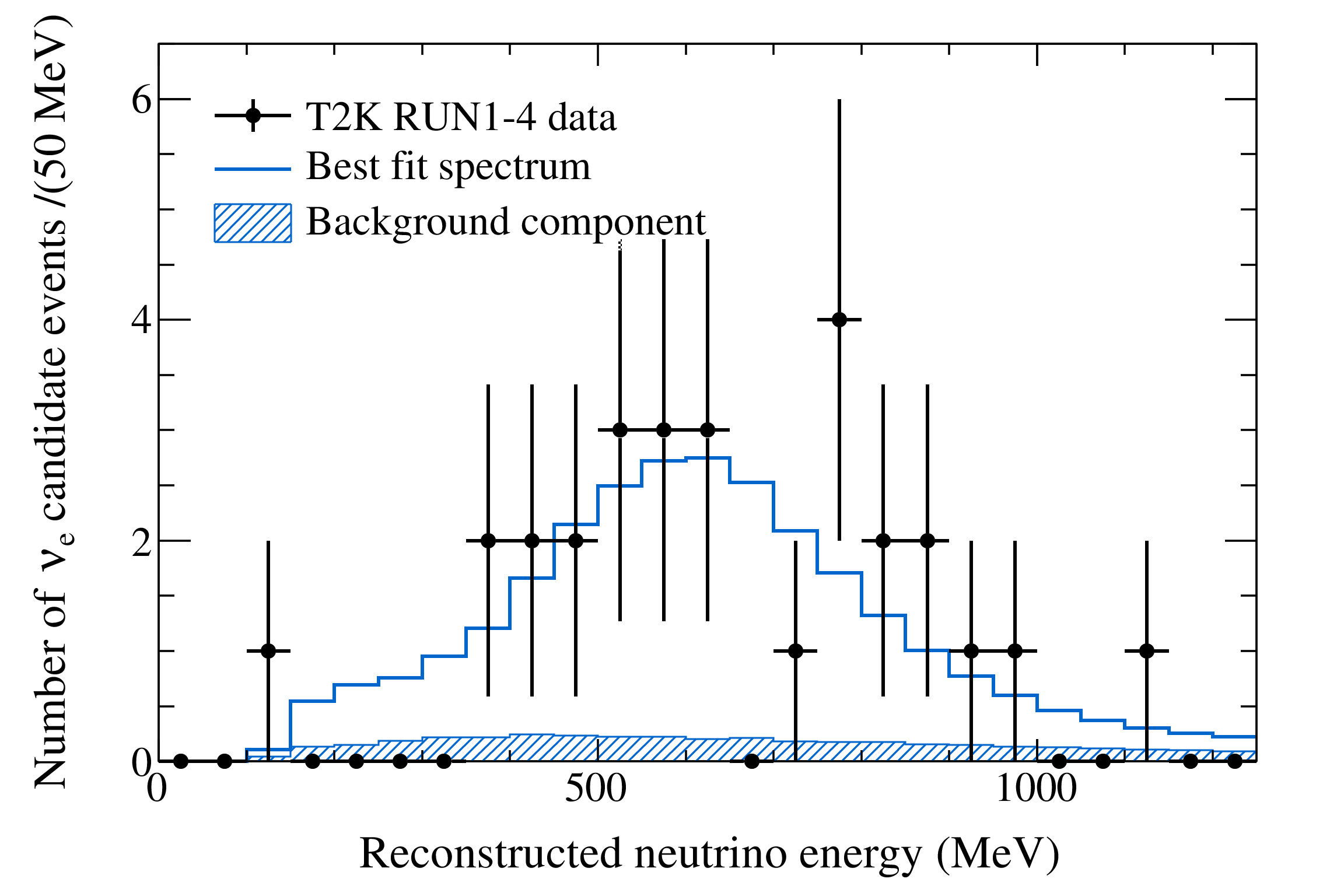} 
 \hspace{5mm}
 \includegraphics[width=0.45\textwidth]{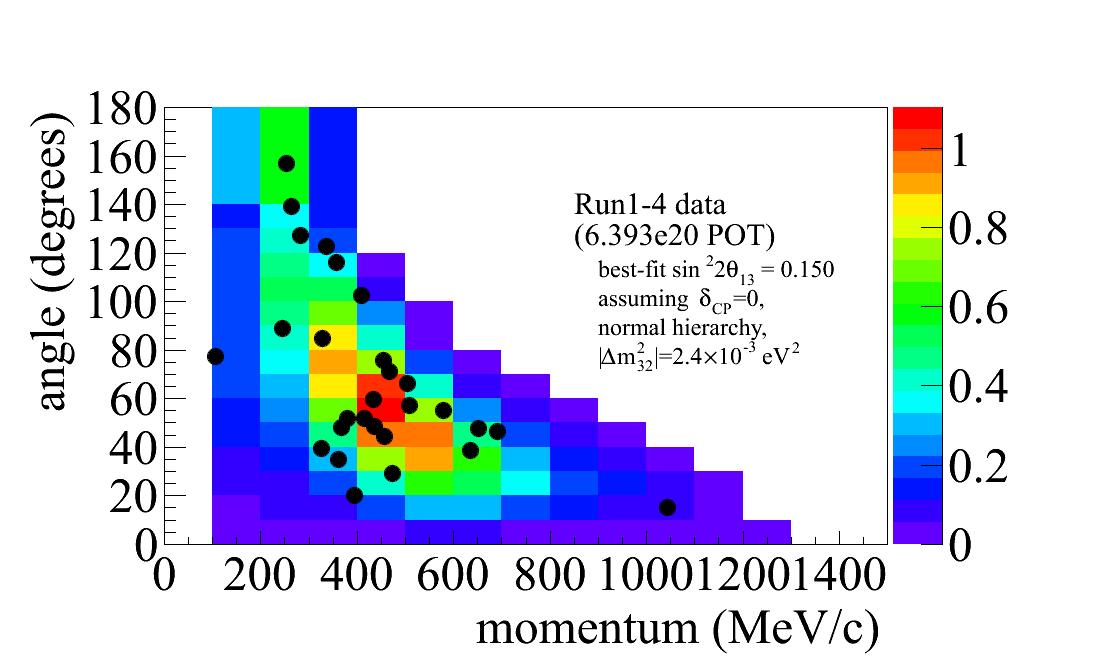}
 \end{center}
 \caption{The final $\nu_e$ candidate sample is shown in both neutrino
   energy (left) and in electron angle relative to the beam direction
   versus the electron momentum.}  \label{fig:finalsamples}
\end{figure}

To extract measurements of the oscillations parameters, the final data
sample is separately fit in both the one-dimensional neutrino energy
distribution and the two-dimensional distribution of electron angle
and electron momentum. Figure~\ref{fig:finalsamples} shows the
selected data sample in both of these distributions compared to the
best-fit MC in each case.

\begin{figure}[h!]
 \begin{center}
 \hspace{6mm}
 \includegraphics[width=0.45\textwidth]{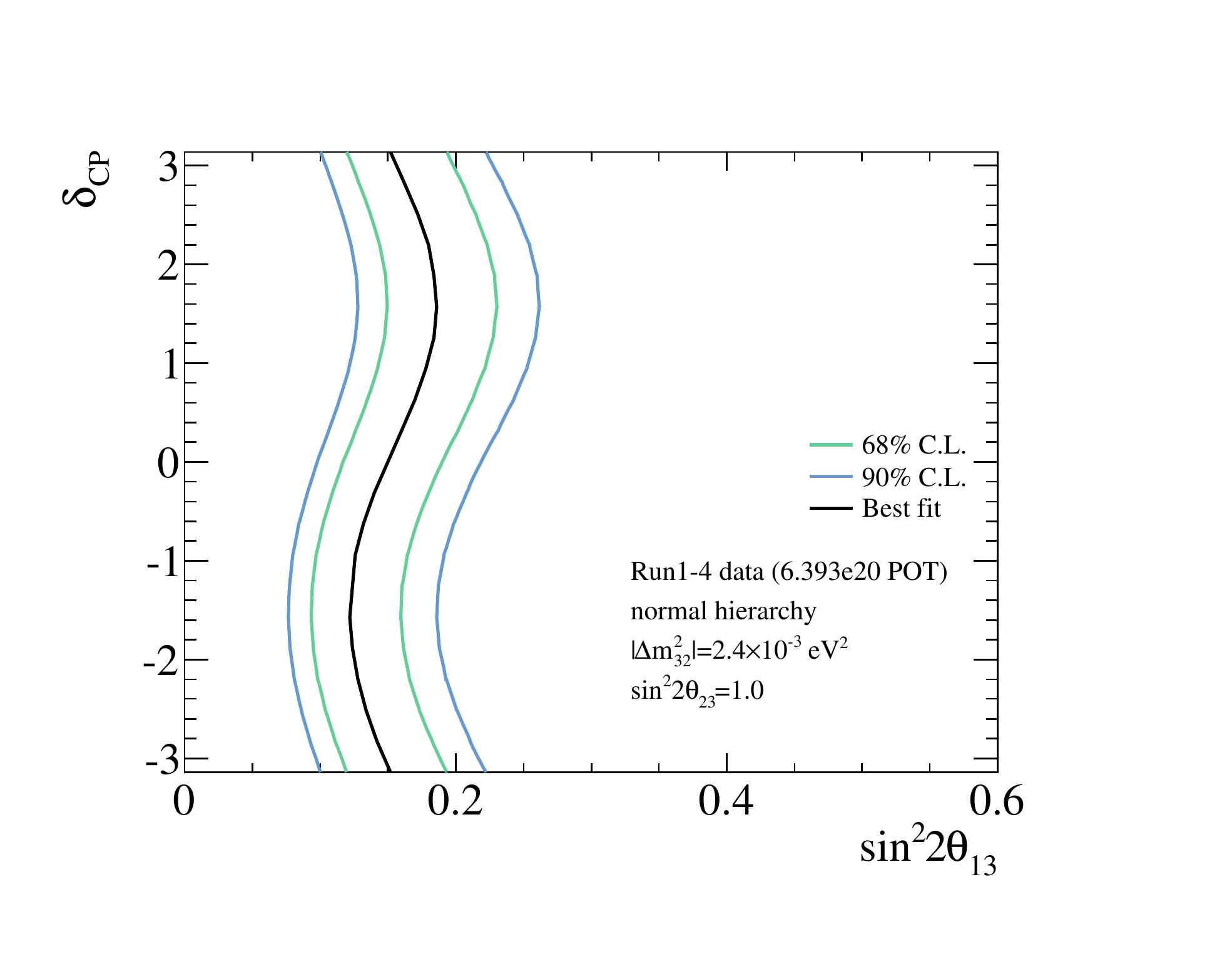} 
 \includegraphics[width=0.45\textwidth]{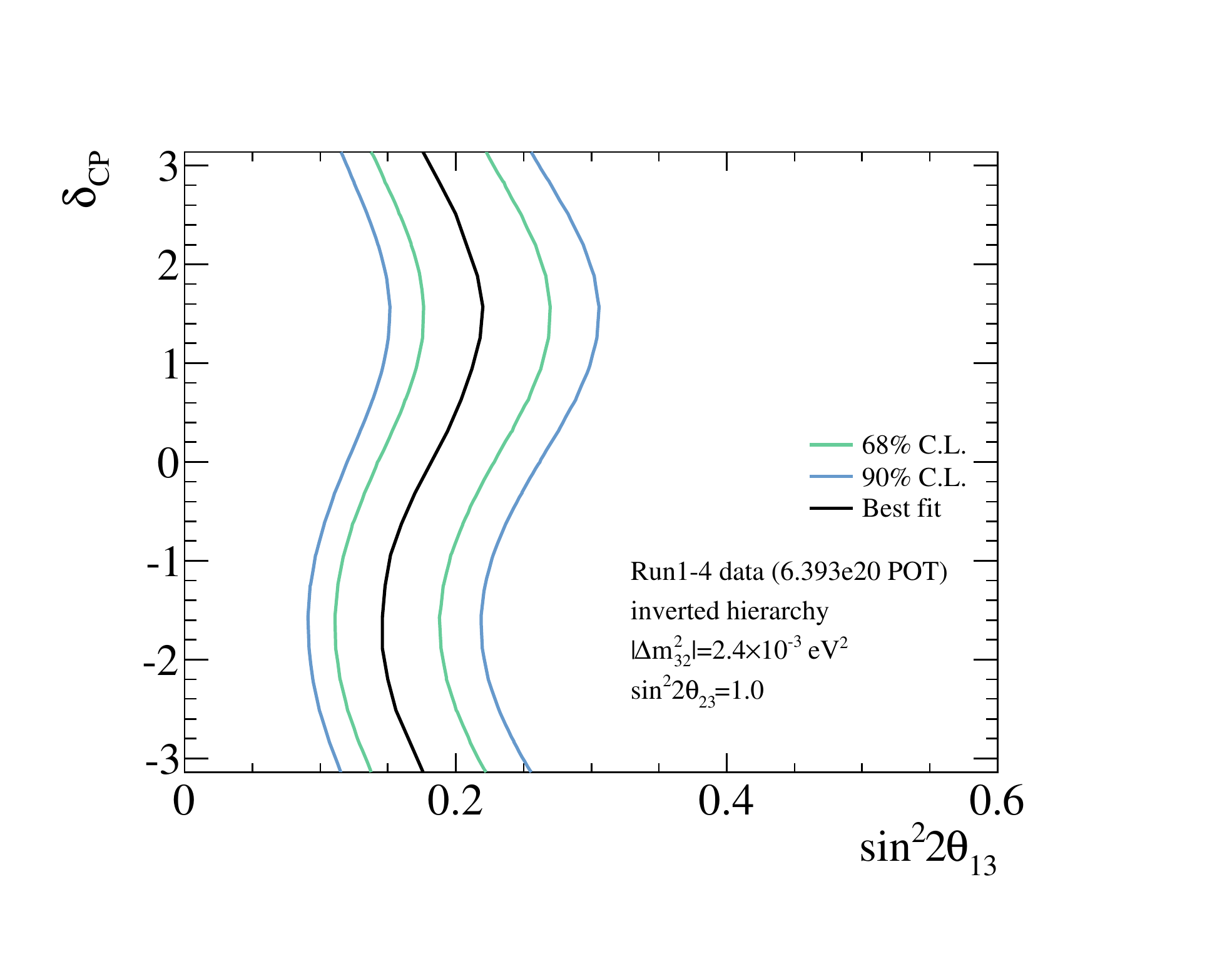}
 \includegraphics[width=0.38\textwidth]{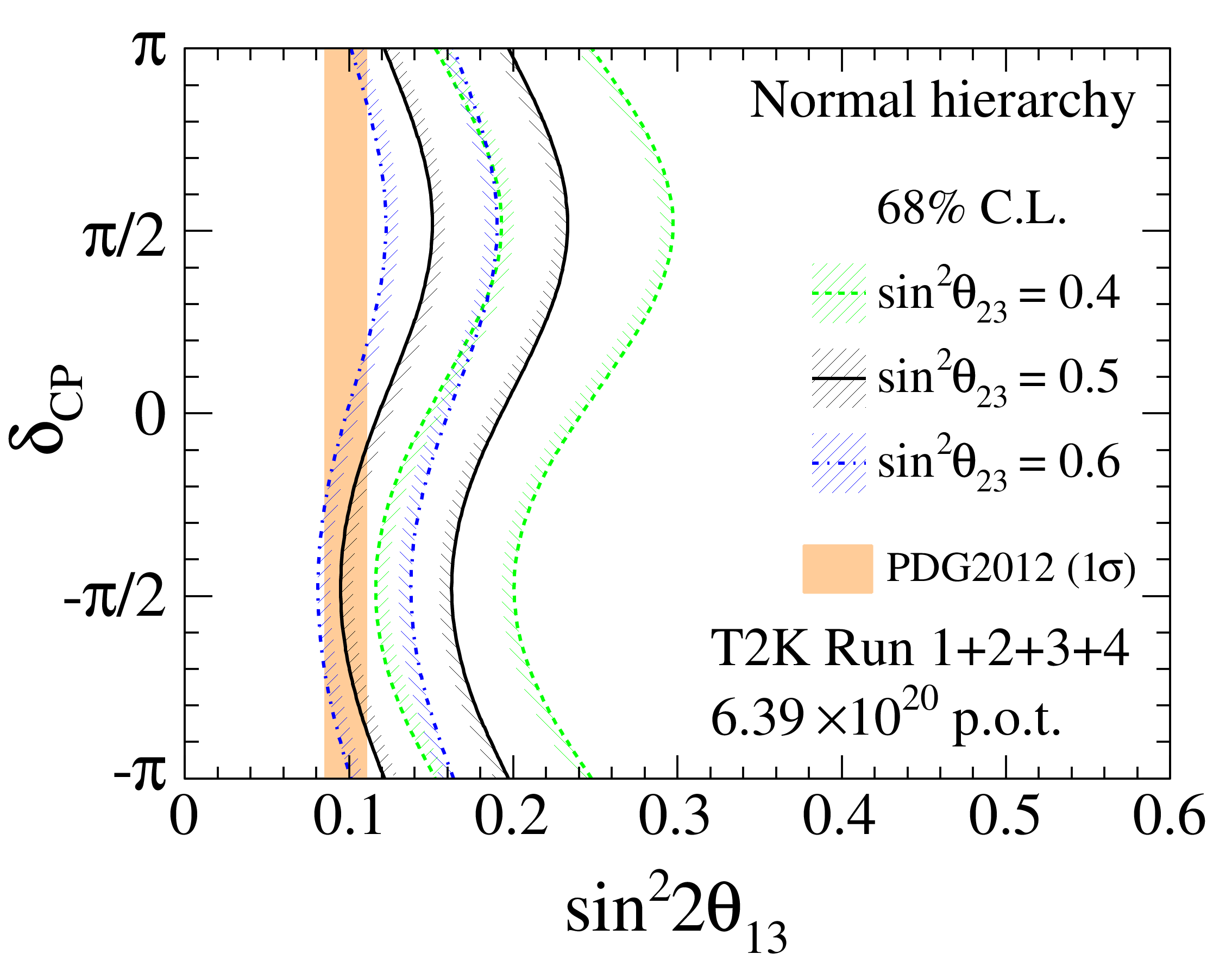} 
 \hspace{10mm}
 \includegraphics[width=0.38\textwidth]{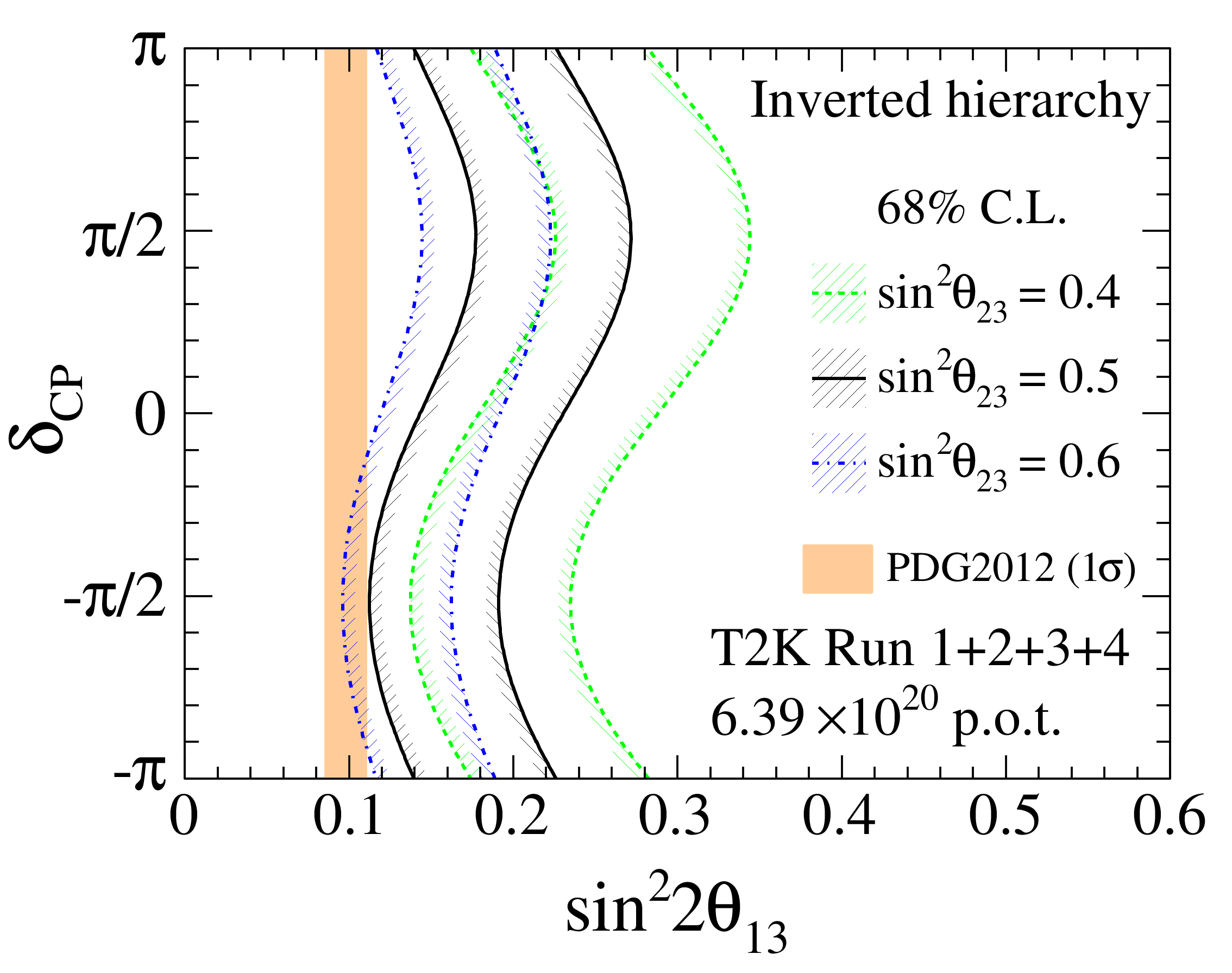} 
 \end{center}
 \caption{The T2K measured 68\% and 90\% confidence level (C.L.)
   contours in sin$^22\theta_{13}$ at many fixed values of
   $\delta_{CP}$, and with all other oscillation parameters fixed, are
   shown assuming normal neutrino mass hierarchy (top-left) and
   inverted hierarchy (top-right), and with sin$^22\theta_{23}$ fixed
   at 0.5. The effect on the 68\% C.L. contours of fixing
   $\theta_{23}$ at near the boundaries of its current 90\%
   C.L. allowed region for both normal hierarchy (bottom-left) and
   inverted hierarchy (bottom-right). The PDG2012~\cite{pdg2012}
   average value of $\theta_{13}$ measured by reactor experiments is
   also shown in the bottom figures.}  \label{fig:theta13contours}
\end{figure}

In the fits presented here $\theta_{13}$ is treated as a free
parameter with all other oscillation parameters fixed to allow for
more straightfoward comparisons with previous T2K results, and to
reduce the dependence of these results on the current uncertainties in
the other oscillation parameters, which are expected to be improved in
the near future. Figure~\ref{fig:theta13contours} shows the allowed
region of sin$^22\theta_{13}$ for various fixed values of
$\delta_{CP}$ with sin$^22\theta_{12}=0.306$, $\Delta
m^2_{21}=7.5\times 10^{-5}~\textrm{eV}^2$~\cite{fogli},
sin$^2\theta_{23}=0.5$, $\left|\Delta m^2_{32}\right|=2.4\times
10^{-3}~\textrm{eV}^2$~\cite{t2knumu}.  If $\delta_{CP}$ is fixed to
zero, the significance of non-zero $\theta_{13}$, calculated via
either the difference of log-likelihood values between the best fit
$\theta_{13}$ and $\theta_{13}=0$ or by generating a large number of
toy experiments assuming $\theta_{13}=0$, is 7.5$\sigma$. The
significance of non-zero $\theta_{13}$ stays above 7$\sigma$ for all
choices of $\delta_{CP}$ and all choices of sin$^22\theta_{23}$ within
its current uncertainty.  The effect of variations in $\theta_{23}$ is
also shown in Figure~\ref{fig:theta13contours} by shifting
sin$^2\theta_{23}$ to values near the currently allowed 90\%
C.L. limits.

\begin{figure}[h!]
 \begin{center}
   \includegraphics[width=0.5\textwidth]{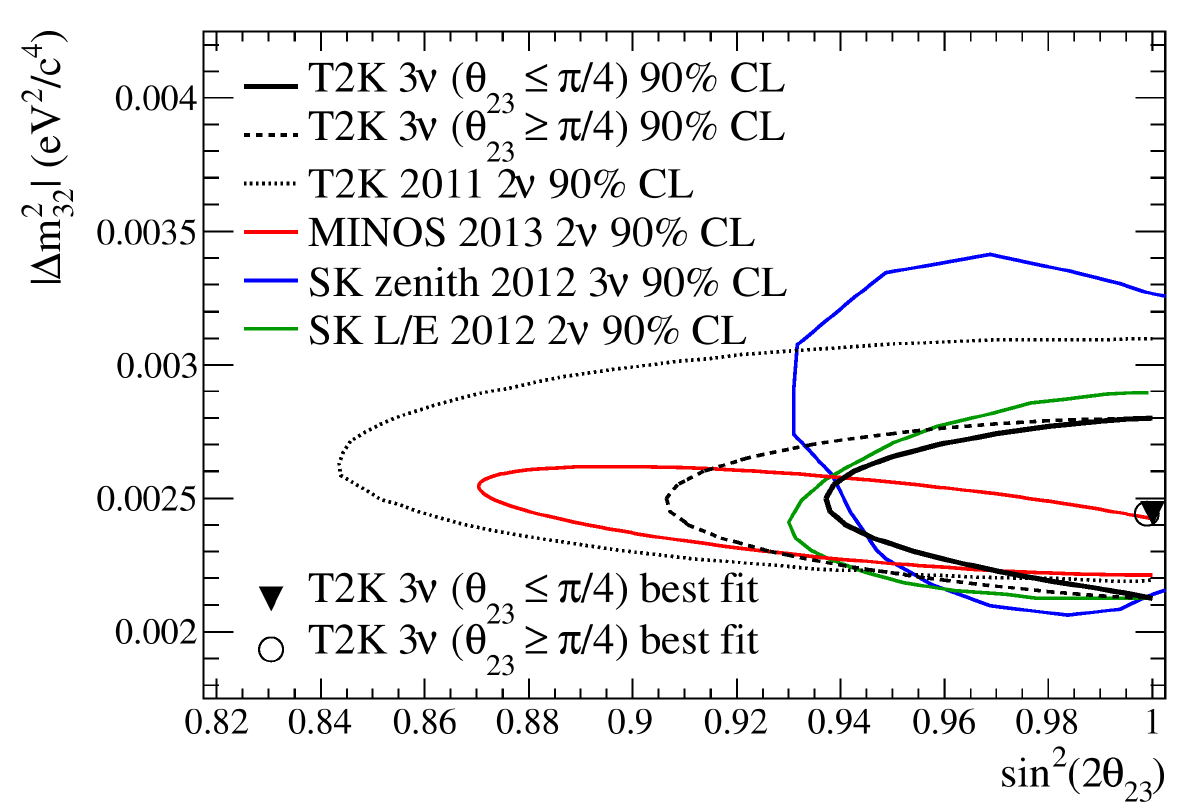} 
 \end{center}
 \caption{Measurements of $\left|\Delta m^2_{32}\right|$ vs
   sin$^22\theta_{23}$ from Super-Kamiokande~\cite{sknumu},
   MINOS~\cite{minosnumu}, and T2K~\cite{t2knumuold} are shown. The
   most recent T2K result is shown separately for the two choices of
   the $\theta_{23}$ octant.}  \label{fig:numudisapp}
\end{figure}

Finally, the T2K $\nu_\mu$-disappearance analysis results have been
updated to better characterize the effect of full 3-flavor neutrino
oscillations. In the previous 2-flavor approximation, the measured
contours in $\left|\Delta m^2_{32}\right|$ vs sin$^22\theta_{23}$ were
independent of whether $\theta_{23}$ was above or below $\pi/2$
radians. However, in a full 3-flavor treatment, particularly now that
$\theta_{13}$ is known to be large, the size of the contour changes
significantly depending on the $\theta_{23}$ octant. To illustrate this effect, Figure~\ref{fig:numudisapp} shows $\left|\Delta m^2_{32}\right|$ vs sin$^22\theta_{23}$ contours have been made separately for $\theta_{23}<\pi/2$ and $\theta_{23}>\pi/2$. In the time since this presentation was given, T2K has published a more complete treatment of this effect in which the best-fit regions are reported in terms of sin$^2\theta_{23}$ rather than sin$^22\theta_{23}$~\cite{t2knumu}.

\end{document}